\documentclass[11pt]{article}
\usepackage{enumerate}
\usepackage{color}
\usepackage{amsmath}
\usepackage{graphicx}
\usepackage{amsmath}
\usepackage{dsfont}
\usepackage{amssymb}
\usepackage{bm}
\title{Black holes with  Lambert W function horizons}
\author{Moises Bravo Gaete\footnote{mbravo-at-ucm.cl
},\,\, Sebastian Gomez\footnote{ sebastian.gomez-at-uautonoma.cl }
\, and Mokhtar~Hassaine\footnote{hassaine-at-inst.mat.utalca.cl}\\
$^{*}$Facultad de Ciencias B\'asicas, Universidad Cat\'olica del
Maule, Casilla 617, Talca, Chile.\\
$^{\dag}$ Facultad de Ingenier\'ia, Universidad Aut\'onoma de Chile, 5 poniente 1670, Talca, Chile.\\
$^{\ddag}$Instituto de Matem\'atica y F\'{\i}sica, Universidad de
Talca, Casilla 747, Talca, Chile.} \textheight 16 cm\textwidth 16 cm
\let\ssection=\section
\renewcommand{\section}{\setcounter{equation}{0}\ssection}
\begin{document}
\maketitle
\begin{abstract}
We consider Einstein gravity with a negative cosmological constant
endowed with distinct matter sources. The different models analyzed
here share the following two properties: (i) they admit static
symmetric solutions with planar base manifold characterized by their
mass and some additional Noetherian charges, and (ii) the
contribution of these latter in the metric has a slower falloff to
zero than the mass term, and this slowness is of logarithmic order.
Under these hypothesis, it is shown that, for suitable bounds
between the mass and the additional Noetherian charges, the
solutions can represent black holes with two horizons whose
locations are given in term of the real branches of the Lambert W
functions. We present various examples of such black hole solutions
with electric, dyonic or axionic charges with AdS and Lifshitz
asymptotics. As an illustrative example, we construct a purely AdS
magnetic black hole in five dimensions with a matter source given by
three different Maxwell invariants.
\end{abstract}

\section{Introduction}
The AdS/CFT correspondence has been proved to be extremely useful
for getting a better understanding of strongly coupled systems by
studying classical gravity, and more specifically black holes. In particular, the
gauge/gravity duality can  be a powerful tool for analyzing finite
temperature systems in presence of a background magnetic field. In such cases, from the
dictionary of the correspondence, the black holes must be
endowed with a magnetic charge corresponding to the external
magnetic field of the CFT. In light of this constatation, it is clear that dyonic black holes are
of great importance in order to study the charge transport at
quantum critical point, particulary for strongly coupled CFTs in
presence of an external magnetic field. For example, four-dimensional dyonic
black holes have been proved to be relevant for a better comprehension of planar condensed matter phenomena
such as the quantum Hall effect \cite{Hartnoll:2007ai},
the superconductivity-superfluidity \cite{Hartnoll:2008vx} or
the Nernst effect \cite{Hartnoll:2007ih}. The study of dyonic black
holes is not only interesting in four dimensions, but also in higher
dimensions where their holographic applications have been discussed
in the current literature. For example, it has been shown that large
dyonic AdS black holes are dual to stationary solutions of a charged
fluid in presence of an external magnetic field
\cite{Caldarelli:2008ze}. In this last reference, the AdS/CFT correspondence was
used conversely and stationary solutions of the Navier-Stokes
equations were constructed corresponding to an hypothetical
five-dimensional AdS dyonic rotating black string with nonvanishing
momentum along the string. We can also mention that magnetic/dyonic
black holes present some interest from a purely gravity point of
view. Indeed, there is a wide range of contexts in which magnetic/dyonic
 solutions are currently studied including in particular
supergravity models \cite{Chamseddine:2000bk, Chow:2013gba},
Einstein-Yang-Mills theory \cite{Nolan:2012ax} or nonlinear
electrodynamics \cite{Bronnikov:2017xrt}. Nevertheless, in spite of
partial results, the problem of finding magnetic solutions in higher
dimension is an highly nontrivial problem. For example, it is easy
to demonstrate that under suitable hypothesis, magnetic solutions in
odd dimensions $D\geq 5$ for the Einstein-Maxwell or for the
Lovelock-Maxwell theories do not exist
\cite{Ortaggio:2007hs,Maeda:2010qz}. This observation is in contrast
with the four-dimensional situation where static dyonic
configuration can be easily constructed thanks to the
electromagnetic duality which rotates the electric field into the
magnetic field. In the same register, one may also suspect the lack
of electromagnetic duality and of the conformal invariance in
dimension $D>4$ to explain the difficulty for constructing the
higher-dimensional extension of the Kerr-Newmann solution.

The purpose of the present paper is twofold. Firstly, we would like
to present a simple dyonic extension of the five-dimensional
Reissner-Nordstrom solution with planar horizon. The solution will
be magnetically charged by considering an electromagnetic source
composed by {\it at least} three different Maxwell gauge fields.
Each of these $U(1)$ gauge fields will be sustained by one of the
three different coordinates of the planar base manifold.
Interestingly enough, the magnetic contribution in the metric has an
asymptotically logarithmic falloff of the form $\frac{\ln r}{r^2}$.
Nevertheless, in spite of this slowly behavior, the thermodynamics analysis yields finite quantities
even for the magnetic charge. Since we are working in five dimensions, we
extend as well this dyonic solution to the case of Einstein-Gauss-Bonnet gravity. We can also mention
that the causal structure of the dyonic solution can not be done analytically. Nevertheless from
different simulations, one can observe that the solution has a Reissner-Nordstrom like behavior. Indeed,
depending on the election of the integration constants, the solution can be a black
hole with inner and outer horizons or an extremal black hole or the
solution can have a naked singularity located at the origin. On the other
hand, we notice that the horizon structure of the purely magnetic solution can be treated
analytically. More precisely, we will show that, as for the
Reissner- Nordstrom solution, the absence of naked singularity can
be guaranteed for a suitable bound relation between the mass and the magnetic
charge. Moreover, in this case, the location of the inner and outer
horizons are expressed analytically in term of the real branches of
so-called Lambert W function. This latter is defined to be the
multivalued inverse of the complex function $f(\omega)=\omega
e^{\omega}$ which  has an infinite countable number of branches but
only two of them are real-valued, see Ref. \cite{LambertW} for a nice review. The
Lambert W functions have a wide range of applications as for example
in combinatoric with the tree functions that are used in the
enumeration of trees \cite{combinatoric} or for equations with delay that
have applications for biological, chemical or physical phenomena,
see e. g. \cite{delayEq} or in the AdS/CFT correspondence as in the expression
of the large-spin expansion of the energy of the Gubser-Klebanov-Polyakov string theory \cite{Floratos:2013cia}.
Just to conclude this parenthesis about the Lambert W function, we also mention that this function
can be used  in the case of the Schwarzschild metric as going from the Eddington-Finkelstein coordinates
to the standard Schwarzschild coordinates

The plan of the paper is organized as follows. In the next section,
we present our toy model for dyonic solutions which consists on the five-dimensional
Einstein-Gauss-Bonnet action with three different Abelian gauge fields. For
this model, we derive a dyonic black hole configuration as well as
its GR limit. A particular attention will be devoted to the purely
magnetic GR solution for which a bound relation between the mass and the
magnetic charge ensures the existence of an event horizon covering
the naked singularity. In this case, the inner and outer
horizons are expressed in term of the two real branches of the
Lambert W functions. We will establish that this mass bound is essentially due to the fact that
the magnetic charge has a slower falloff of logarithmic order to
zero than the mass term in the metric function. Starting from this
observation, we will present in Sec. $3$ various examples of black holes sharing
this same feature with electric, axionic or magnetic charges and
with AdS and Lifshitz asymptotics. In Sec. $4$, we extend
the previous solutions to general dyonic configurations with axionic charges. Finally, the last
section is devoted to our conclusion and an appendix is provided
where some useful properties of the Lambert W functions are given.

\section{Five-dimensional dyonic black hole solution}
In Refs.  \cite{Ortaggio:2007hs,Maeda:2010qz}, it has been proved
that, under suitable hypothesis, magnetic black hole solutions for
Einstein-Maxwell action in odd dimensions $D\geq 5$ can not exit. As we will show below, a simple
way of circumventing this obstruction is to consider more than one
Maxwell gauge  field. The fact of considering various Abelian fields in order to construct
dyonic black holes in five dimensions have already been considered, see Refs. \cite{Arias:2017yqj} and \cite{Bravo-Gaete:2017nkp}.
More precisely, we will establish that the
Einstein gravity eventually supplemented by the Gauss-Bonnet term
since we are working in $D=5$ can admit dyonic black hole solutions
for an electromagnetic source given at least by three different
Maxwell invariants. In order to achieve this task, we consider the following action
\begin{eqnarray}
S[g, {\cal A}_I]=\int d^{5}x\,\sqrt{-g} \Big[\frac{R-2 \Lambda}{2}
+\frac{\alpha}{2} \left(R^{2}-4\,R_{\mu \nu}R^{\mu
\nu}+R_{\alpha\beta\mu\nu}R^{\alpha\beta\mu\nu}\right)-\frac{1}{4}\sum_{I=1}^{3}
{\cal{F}}_{(I) \mu \nu}{\cal{F}}_{(I)}^{\mu \nu}\Big], \label{Ik2}
\end{eqnarray}
where the ${\cal{F}}_{(I) \mu \nu}$'s  are the three different
Maxwell field strengths associated to the $U(1)$ gauge fields ${\cal
A}_I$ for $I=\{1, 2, 3\}$ and $\alpha$ represents the
Gauss-Bonnet coupling constant. The field equations obtained by
varying this action read
\begin{eqnarray}
\label{fieldeqs} &&G_{\mu \nu}+\Lambda g_{\mu \nu}+\alpha
K^{\tiny{\mbox{GB}}}_{\mu \nu}
=\sum_{I=1}^{3}\left({\cal{F}}_{(I)\mu \sigma} {\cal{F}}_{(I)
\nu}^{\phantom{\sigma \sigma \, } \sigma}\right)-\frac{1}{4}
{g}_{\mu \nu}\sum_{I=1}^{3}\left(
        {\cal{F}}_{(I) \sigma \rho}{\cal{F}}_{(I)}^{\sigma
\rho}\right),\nonumber\\
&&{\nabla}_{\mu}{\cal{F}}_{(I)}^{\mu \nu}=0,\qquad \mbox{for}\quad
I=\{1,2, 3\},
\end{eqnarray}
where  the variation of the Gauss-Bonnet term is given by
\begin{eqnarray*}\label{tensorgb}
K^{\tiny{\mbox{GB}}}_{\mu \nu}=
2\big(RR_{\mu\nu}-2R_{\mu\rho}R^{\rho}_{\,\,\nu}-2R^{\rho\sigma}R_{\mu\rho\nu\sigma}
+R_{\mu}^{\,\,\rho\sigma\gamma}R_{\nu\rho\sigma\gamma}\big)
-\frac{1}{2}\, g_{\mu\nu}\big(R^{2}-4 R_{\rho \sigma} R^{\rho
\sigma} +R_{\rho \sigma \lambda \delta}R^{\rho \sigma \lambda
\delta}\big).
\end{eqnarray*}

In one hand, it is known that the field equations (\ref{fieldeqs})
with one Maxwell invariant $I=1$ admits electrically charged black
holes \cite{Cvetic:2001bk} generalizing the solution of
Boulware-Deser \cite{Boulware:1985wk}. On the other hand, it is simple to prove that the magnetic
extension of the  Boulware-Deser solution can
not exist \cite{Ortaggio:2007hs,Maeda:2010qz}. Nevertheless, as shown below, the presence
of two extra Maxwell invariants renders possible the magnetic
extension of the Boulware-Deser solution but only in the case of
flat horizon. In fact a dyonic solution with flat horizon of the
field equations (\ref{fieldeqs}) is found to be
\begin{eqnarray}
\label{solution} &&ds^2=-F(r)dt^2+\frac{dr^{2}}{F(r)}+ r^{2}
\sum_{i=1}^{3}
dx_{i}^{2},\nonumber\\
&&F(r)=\frac{r^2}{4 \alpha}\left[ 1\mp \sqrt{1+\frac{4 \alpha
\Lambda}{3}+{\frac {16 \alpha {\cal{M}} }{3 \vert\Sigma_3\vert {r}^{4}}}-{\frac {4 \alpha {\cal{Q}}_{e}^{2}}{ \vert\Sigma_3\vert^2
{r}^{6}}}+\frac{8 \alpha {\cal{Q}}_{m}^2 \ln(r)}{\vert\Sigma_3\vert^2 r^{4}}}\right],\\
&&{\cal A}_1=-\frac{{\cal{Q}}_{e}}{2 \vert\Sigma_3\vert r^2}dt+\frac{{\cal{Q}}_{m}}{\vert\Sigma_3\vert} x_2\, dx_3,\quad {\cal
A}_2=-\frac{{\cal{Q}}_{e}}{2 \vert\Sigma_3\vert r^2}dt+\frac{{\cal{Q}}_{m}}{\vert\Sigma_3\vert} x_3\,dx_1,\nonumber\\
&&{\cal A}_3=-\frac{{\cal{Q}}_{e}}{2 \vert\Sigma_3\vert r^2} dt+\frac{{\cal{Q}}_{m}}{\vert\Sigma_3\vert} x_1\,dx_2,\nonumber
\end{eqnarray}
where ${\cal M}$, ${\cal Q}_{e}$ and ${\cal Q}_{m}$ are three integration constants corresponding respectively  to the mass,
the electric and the magnetic charge and $\vert\Sigma_3\vert$ is the finite volume of the compact
$3-$dimensional flat base manifold. Various comments can
be made concerning this dyonic solution. Firstly, in the absence of
the magnetic charge {${\cal Q}_{m}=0$}, the solution reduces to the electrically
extension of the Boulware-Deser solution \cite{Cvetic:2001bk} even
if there are three different Maxwell invariants. This is because
each of these three invariants contributes in the same footing for
the full solution, and hence one could have switch off two of them
from the very beginning. The GR limit $\alpha\to 0$ of the solution
concerns only the upper branch of the solution and yields to the
metric function given by
 \begin{eqnarray}
\label{GRsol} F_{\mbox{{\tiny
GR}}}(r)=-\frac{r^2\Lambda}{6}-\frac{2 {\cal M}}{3 \vert\Sigma_3\vert r^2}+\frac{{\cal Q}_{e}^2}{2 \vert\Sigma_3\vert^2
r^4}-\frac{ {\cal Q}_{m}^2\ln r}{\vert\Sigma_3\vert^2 r^2},
\end{eqnarray}
while the Abelian gauge fields remain identical. Computing the
Kretschmann invariant, one notices that the dyonic solution  in the
Einstein-Gauss-Bonnet theory or its GR limit has a singularity
located at the origin.

The causal structure of the dyonic solution is quite involved and can not
be treated analytically as in the case of the four-dimensional
Reissner-Nordstrom dyonic solution. Nevertheless, it is quite simple to see that the GR
solution (\ref{GRsol}) with $\Lambda<0$ and without magnetic charge,
has a Reissner-Nordstrom like behavior in the sense
that for {${\cal M}\geq { 3^{\frac{5}{3}} \vert {\cal
Q}_{e}\vert^{\frac{4}{3}}\left(-\Lambda\right)^{\frac{1}{3}}}\big{/}
{4^{\frac{4}{3}} \vert \Sigma_3 \vert ^{\frac{1}{3}}}$, the solution
describes a (extremal) black hole while the case { ${\cal M}<{
3^{\frac{5}{3}} \vert {\cal
Q}_{e}\vert^{\frac{4}{3}}\left(-\Lambda\right)^{\frac{1}{3}}}\big{/}
{4^{\frac{4}{3}} \vert \Sigma_3 \vert ^{\frac{1}{3}}}$} will yield a
naked singularity. The dyonic GR solution has also a similar behavior which can
be appreciated only by means of some simulations reported in the graphics below.
In the next subsection, we will see that in the purely magnetic case, the causal structure
of the solution can be analyzed analytically.

To conclude this section, we mention that the GR dyonic solution
(\ref{GRsol}) satisfies the dominant energy conditions. Indeed, it
is simple to see that the energy density $\mu$, the radial pressure
$p_r$ and the tangential pressure $p_t$ given by
\begin{eqnarray}
\mu=\frac{3}{2\vert\Sigma_3\vert^2 r^6}\left[{\cal Q}_m^2r^2+{\cal
Q}_e^2\right],\qquad p_r=-\mu,\qquad
p_t=\frac{1}{2\vert\Sigma_3\vert^2 r^6}\left[{\cal Q}_m^2r^2+3{\cal
Q}_e^2\right], \label{DECdyonic}
\end{eqnarray}
verify the dominant energy conditions
\begin{eqnarray}
\mu\geq 0,\qquad -\mu\leq p_r\leq \mu,\qquad -\mu\leq p_t\leq \mu.
\label{DEC}
\end{eqnarray}

\begin{figure}[h!]
\centering
\begin{tabular}{cc}
\includegraphics[width=0.45\textwidth]{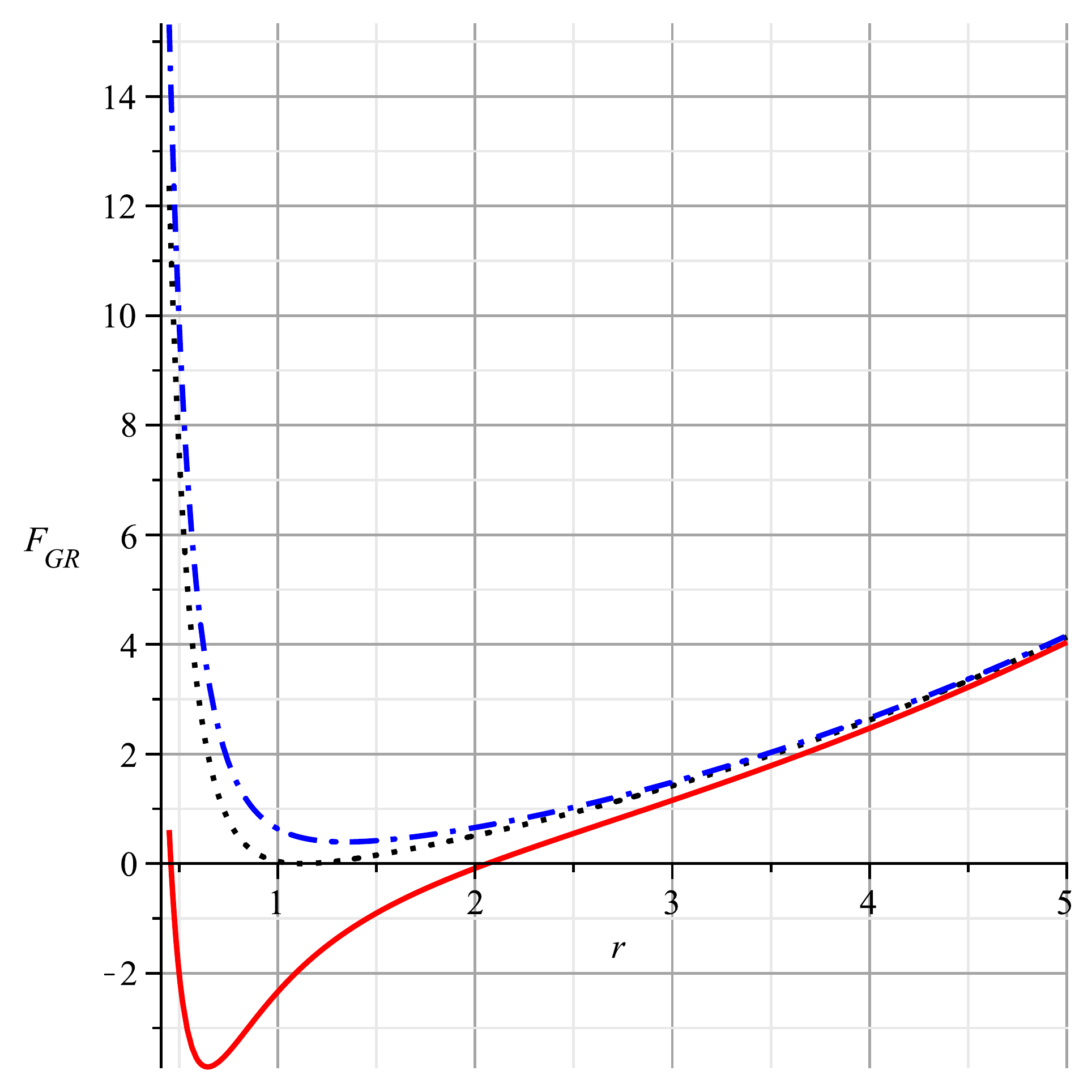} &  \includegraphics[width=0.45\textwidth]{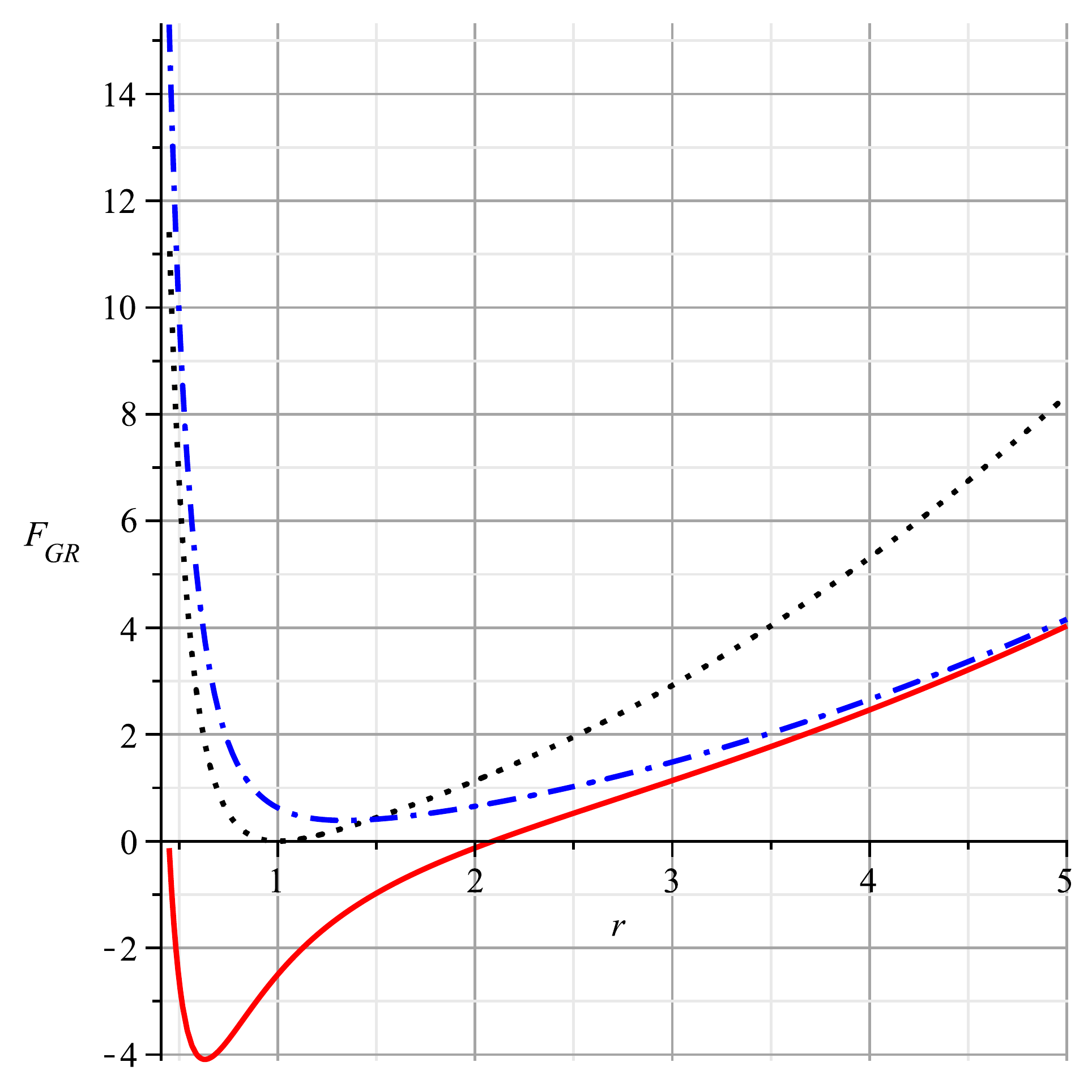}
\end{tabular}
\caption{Plot of the metric function $F(r)$ in the GR-limit
(\ref{GRsol}), where the left panel corresponds to the electric
solution for $\Lambda=-1$ and {${\cal Q}_{e}/\vert \Sigma_3 \vert=2/\sqrt{3}$, while the right panel is the
dyonic situation with the same values of $\Lambda$ and ${\cal Q}_{e}/\vert \Sigma_3 \vert$, together
with ${\cal Q}_{m}/\vert \Sigma_3 \vert =0.1$}. For both cases, the naked singularity solution is
represented by a blue dashed-dotted line, the extremal black holes
with a black dashed line and the solution with inner and outer
horizons by a red continuous line. } \label{fig1}
\end{figure}

\subsection{Purely magnetic GR solution}
For $ \alpha\to 0$ and ${\cal Q}_e=0$, the purely magnetic GR solution (\ref{GRsol}) becomes
\begin{eqnarray}
\label{magnsolution} &&ds^2=-F(r)dt^2+\frac{dr^{2}}{F(r)}+ r^{2}
\sum_{i=1}^{3} dx_{i}^{2},\qquad
F(r)=-\frac{r^2\Lambda}{6}-\frac{2{\cal M}}{3\vert\Sigma_3\vert
r^2}-\frac{{\cal Q}_m^2\ln r}{\vert\Sigma_3\vert^2 r^2},\nonumber\\ &&{\cal
A}_1=\frac{{\cal Q}_m}{\vert\Sigma_3\vert}x_2\, dx_3,\quad {\cal
A}_2=\frac{{\cal Q}_m}{\vert\Sigma_3\vert} x_3\,dx_1,\quad \quad
{\cal A}_3=\frac{{\cal Q}_m}{\vert\Sigma_3\vert} x_1\,dx_2.
\end{eqnarray}
In order to study the variations
of the metric function $F$, it is useful to define
\begin{eqnarray}
h(x)=6 \vert\Sigma_3\vert^2 r^2 F(r)=-\Lambda \vert\Sigma_3\vert^2
x-\frac{3}{2}{\cal Q}_m^2 \ln x-4{\cal M}\vert\Sigma_3\vert,\qquad
\mbox{with}\quad x=r^4. \label{rel}
\end{eqnarray}
For negative cosmological constant $\Lambda<0$, we have
$\lim_{x\to\infty}h(x)=\lim_{x\to 0^+}h(x)=\infty$ and the function
$h$ has a global minimum at $x=\frac{3{\cal
Q}_m^2}{-2\Lambda\vert\Sigma_3\vert^2}$. The equation for the zeros
of the function $h$ that will give as well the location of the
horizons for the metric function $F$ through (\ref{rel}) is of the form
(\ref{eqLW1}). Hence, its corresponding discriminant as defined in Eq.
(\ref{discriminant}) is given by
\begin{eqnarray}
\Delta=\frac{2\Lambda \vert\Sigma_3\vert^2}{3{\cal
Q}_m^2}e^{-\frac{8{\cal M} \vert\Sigma_3\vert}{3{\cal Q}_m^2}}.
\label{deltaaa}
\end{eqnarray}
Since we are considering the negative cosmological constant case $\Lambda<0$, the discriminant is
negative, and as mentioned in the appendix, the equation $h(x)=0$ will
have two real roots only if $\Delta\in ]-\frac{1}{e},0[$. This condition in
turn requires that the mass ${\cal M}$ must satisfy the following
bound relation
\begin{eqnarray}
{\cal M}>\frac{3{\cal Q}_m^2}{8\vert\Sigma_3\vert}
\left[1-\ln\left(\frac{3{\cal Q}_m^2}{-2\Lambda
\vert\Sigma_3\vert^2}\right)\right]={\cal M}_0.\label{boundmass}
\end{eqnarray}
For ${\cal M}$ satisfying such bound, the metric
function $F$ has an inner (Cauchy) horizon $r_{-}$ and an outer
(event) horizon $r_{+}$ whose locations are expressed in term of the
two real branches of the Lambert W functions, $W_{0}$ and $W_{-1}$
as
\begin{eqnarray}
r_{-}=e^{-\frac{W_{0}(\Delta)}{4}-\frac{2{\cal
M}\vert\Sigma_3\vert}{3{\cal Q}_m^2}},\qquad\qquad
r_{+}=e^{-\frac{W_{-1}(\Delta)}{4}-\frac{2{\cal
M}\vert\Sigma_3\vert}{3{\cal Q}_m^2}},
\end{eqnarray}
with $\Delta$ given  by (\ref{deltaaa}). In contrast with the
four-dimensional magnetic Reissner-Nordstrom solution (or even the
dyonic configuration), the bound (\ref{boundmass}) does not restrict
the mass ${\cal M}$ to be positive. In fact, for ${\cal Q}_m^2\geq
-2\Lambda \vert\Sigma_3\vert^2 e/3$, the bound ${\cal M}_0\leq 0$,
and hence the singularity at the origin can still be covered by an
horizon even for a solution with a negative mass. On the other hand, for ${\cal M}$
saturating the  bound (\ref{boundmass}), namely ${\cal M}={\cal
M}_0$ or equivalently $\Delta=-\frac{1}{e}$, one ends up with an
extremal black hole with $r_{+}=r_{-}$. Finally, for ${\cal M}<{\cal
M}_0$, the solution will have a naked singularity. To be complete, we also mention
that the energy density, the radial and tangential pressure of the purely magnetic GR solution
are given by (\ref{DECdyonic}) with ${\cal Q}_e=0$, and hence the magnetic
solution satisfies as well the dominant energy conditions
(\ref{DEC}).

\section{Other examples of black holes with  Lambert W function horizons}
In the previous section, we have shown that purely magnetic black
holes of five-dimensional Einstein gravity with $3$ different
Abelian gauge fields exist provided a certain bound relation between
the mass and the magnetic charge. In addition, the location of the
horizons can be expressed thanks to the real branches of the Lambert
W functions. In this section, we will present few examples enjoying
these same features (bound for the mass and horizons expressed in
term of the Lambert W functions) with different Noetherian charges
(electric, dyonic, magnetic or axionic) and different asymptotics
(AdS or Lifshitz). In order to achieve this task, it is clear from
the previous analysis that the Noetherian charges in the metric must
have a slower falloff of logarithmic order in comparison to the mass
term. In what follows, we will present four different such solutions:
an AdS dyonic black hole in five dimensions,
an AdS electrically charged solution in odd dimension and two
Lifshitz black holes with a magnetic and axionic charge in arbitrary
dimension. These configurations are particular solutions of the following
general $D-$dimensional action
\begin{eqnarray}
\label{Lagdila} &&S[g, \phi, A, {\cal A}, \psi_j]=\int
d^Dx\sqrt{-g}\,{\cal L}, \\&&{\cal L}=\frac{R-2 \Lambda}{2}-
\frac{1}{2}\partial_{\mu}\phi\partial^{\mu}\phi-\frac{1}{4}
e^{\lambda \phi}\left(F_{ \mu \nu}F^{\mu \nu}\right)^q-
 \frac{1}{4}\sum_{I=1}^{n} e^{\alpha_I\phi}{\cal{F}}_{(I) \mu \nu}{\cal{F}}_{(I)}^{\mu
\nu}-\frac{1}{2}\sum_{j=1}^{D-2}e^{\beta_j\phi}
\partial_{\mu} \psi_{j}
\partial^{\mu} \psi_{j}. \nonumber
\end{eqnarray}
In this action, we leave open the possibility of having a nonlinear
Maxwell term $(F_{\mu \nu}F^{\mu \nu})^q$ where $F_{\mu
\nu}=\partial_{\mu} A_{\nu}-\partial_{\nu} A_{\mu}$. Such nonlinearity
has been shown to be fruitful to obtain
charged solutions in different gravity contexts, see e. g.
\cite{nonlineElectro}. As in the previous example, in order to
sustain a magnetic charge, we will also add some extra Abelian gauge
fields ${\cal{F}}_{(I)\mu \nu}=\partial_{\mu}
{\cal{A}}_{(I)\nu}-\partial_{\nu} {\cal{A}}_{(I)\mu}$ for
$I=\{1,2,\cdots,n\}$. We justify the presence of axionic fields
$\psi_j$ from the fact that we are looking for solutions with planar
base manifold, and as shown in the fourth example or in the next
section, the axionic fields perfectly accommodate an ansatz of the
form $\psi_j=\lambda x_j$ where the $x_j$ are the planar coordinates
of the base manifold (\ref{lineelementadm}). This particular ansatz
for the axionic fields also provides a simple mechanism of momentum dissipation
\cite{Andrade:2013gsa}, and in this case the holographic DC conductivities (electrical, thermoelectric
and thermal) can be expressed in terms of the black hole horizon
data \cite{Donos:2013eha, Donos:2014cya}. For examples, the DC
conductivities of dyonic black holes with axionic fields have been
computed recently in Refs. \cite{axiondyonic}. Finally, we note that
the model considered here (\ref{Lagdila}) also allows a possible
coupling of the electromagnetic fields $A$, ${\cal A}$ and the
axionic fields $\psi_j$ to a dilaton field $\phi$.

The field equations associated to the action (\ref{Lagdila}) read {
\begin{eqnarray*}
        G_{\mu \nu}+\Lambda g_{\mu \nu}=\Big(\partial_{\mu} \phi
        \partial_{\nu} \phi-\frac{1}{2} g_{\mu \nu}
        \partial_{\sigma}\phi \partial^{\sigma}\phi\Big)+
        q\, e^{\lambda \phi} \big(F_{\sigma \rho} F^{\sigma \rho}\big)^{q-1}
        F_{\mu \sigma} F_{\nu}^{\phantom{\,} \sigma}
        -\frac{g_{\mu \nu}}{4}
        e^{\lambda
        \phi} \left(F_{\sigma \rho}F^{\sigma \rho}\right)^q&&\nonumber\\
        +\sum_{I=1}^{n}e^{\alpha_{I} \phi} \left({\cal{F}}_{(I)\mu \sigma} {\cal{F}}_{(I) \nu}^{\phantom{\sigma \sigma} \sigma}-\frac{1}{4} {g}_{\mu \nu}{\cal{F}}_{(I) \sigma \rho}{\cal{F}}_{(I)}^{\sigma
        \rho}\right)+\sum_{j=1}^{D-2}e^{\beta_{j} \phi}\Big(\partial_{\mu} \psi_{j}\partial_{\nu} \psi_{j}-\frac{1}{2} g_{\mu \nu}
        \partial_{\sigma}\psi_{j} \partial^{\sigma}\psi_{j}\Big),&&\\
       \nabla_{\mu} \left(e^{\lambda \phi} \big(F_{\sigma \rho} F^{\sigma \rho}\big)^{q-1} F^{\mu
        \nu}\right)=0,&&\\
        {\nabla}_{\mu} \left(e^{\alpha_{I} \phi} {\cal{F}}_{(I)}^{\mu \nu}\right)=0, &&
        \\
        \nabla_{\mu} \left(e^{\beta_{j} \phi} \nabla^{\mu} \psi_{j}\right)=0,&&
        \\
        \Box \phi- \frac{\lambda}{4} e^{\lambda \phi}
         \left(F_{(i) \sigma \rho}
        F_{(i)}^{\sigma
        \rho}\right)^q-\frac{1}{4}\sum_{I=1}^{n} \alpha_{I} e^{\alpha_{I}\phi}{\cal{F}}_{(I) \mu \nu}{\cal{F}}_{(I)}^{\mu
\nu}-\frac{1}{2}\sum_{j=1}^{D-2} \beta_j e^{\beta_j \phi}\partial_{\mu} \psi_{j} \partial^{\mu} \psi_{j}=0,&&
    \end{eqnarray*}}
and we look for a static ansatz with a planar base
manifold of the form
\begin{eqnarray}
ds^2&=&-N^{2}(r)F(r)dt^2+\frac{dr^{2}}{F(r)}+ r^{2} \sum_{i=1}^{D-2}
dx_{i}^{2}. \label{lineelementadm}
\end{eqnarray}
In what follows, we will derive four classes of solutions of the previous
field equations, and their analysis will only be considered in the case of a negative
cosmological constant $\Lambda<0$.

\subsection{Electrically charged AdS black holes for nonlinear Maxwell theory in odd dimension}
This case will correspond of setting $\phi={\cal{A}}_{I}=\psi_j=0$
in (\ref{Lagdila}) and the Maxwell nonlinearity $q$ is of the form
$q=\frac{D-1}{2}$. As shown in Ref. \cite{Maeda:2008ha}, there
exists a purely electric solution with logarithmic falloff, and this
solution, in order to be real, must be restricted to odd dimension
$D=2k+1$ with $k\geq 1$. Hence the Maxwell nonlinearity is
$q=k$ and the metric function and the electric potential are given
by
\begin{eqnarray}
 F(r)&=&-\frac{r^2\Lambda}{k(2k-1)}-\frac{2 {\cal{M}}}{
(2k-1)\vert\Sigma_{2k-1}\vert r^{2k-2}}-(-2)^{k-1} \left[\frac{(-1)^k (2)^{1-k}  {\cal{Q}}_{e}}{k \vert\Sigma_{2k-1}\vert}\right]^{\frac{2k}{2k-1}}
\frac{ \ln r}{r^{2k-2}},\nonumber\\
N(r)&=&1,\quad
A_0=- \left[\frac{ (-1)^k (2)^{1-k}{\cal{Q}}_{e}}{k \vert\Sigma_{2k-1}\vert}\right]^{\frac{1}{2k-1}} \ln r.
\end{eqnarray}
Here ${\cal{M}}$ is the mass, ${\cal{Q}}_{e}$ is the electric charge and $\vert\Sigma_{2k-1}\vert$
denotes the finite volume element of the compact $(2k-1)-$dimensional
base manifold.

As for the previous magnetic solution, the equation determining the
zeros of the metric function $F$ can be put in the form
(\ref{eqLW1}) by substituting $x=r^{2k}$ and in this case, the
discriminant (\ref{discriminant}) is given by
{\begin{eqnarray} \Delta=-\frac{ \vert\Sigma_{2k-1}\vert \Lambda\,{{e}^{{\frac {2 {\cal M} k}{B}}}}}{B}, \qquad \mbox {with} \qquad B =\left( -2 \right) ^{k-2} \left( 2\,k-1 \right)\left[\frac{(-1)^k (2)^{1-k}  {\cal{Q}}_{e}}{k \vert\Sigma_{2k-1}\vert}\right]^{\frac{2k}{2k-1}} \vert\Sigma_{2k-1}\vert.
\end{eqnarray}}
We note that, because of the presence of the term $\left( -2 \right)
^{k-2}$, the sign of the discriminant will depend on the parity of
the integer $k$. Indeed, for even $k$ or equivalently for odd
dimensions $D=5\,\,\mbox{mod}\,\, 4$, the discriminant is positive,
and hence the solution is a black hole for any value of the mass {
${\cal M}$}, and there is a single horizon located at
$$
r_h= e^{-\frac{W_{0}(\Delta)}{2k}-\frac{{\cal M}}{B}}.
$$
Nevertheless, in this case, it is simple to see that the energy
density is always negative, and consequently the energy conditions do not
hold. On the other hand, for odd $k$ or equivalently for odd
dimensions $D=3\,\,\mbox{mod}\,\,4$, the solution will be a black
hole provided that the mass satisfies the following bound relation
with the electric charge
\begin{eqnarray}
{\cal M}\geq { -\frac{B}{2k} \left[1-\ln\left(\frac{B}{ \Lambda
\vert\Sigma_{2k-1}\vert}\right)\right]}. \label{bound1}
\end{eqnarray}
In this case, the inner and outer horizons are given by
\begin{eqnarray}
{ r_{-}=  e^{-\frac{W_{0}(\Delta)}{2k}-\frac{{\cal M}}{B}},\qquad
r_{+}=  e^{-\frac{W_{-1}(\Delta)}{2k}-\frac{{\cal M}}{B}}},
\end{eqnarray}
and the dominant energy conditions (\ref{DEC}) are satisfied with
{\begin{eqnarray}
\mu=\frac{2^{k-2} (2k-1)}{r^{2k}}
\left[\frac{(-1)^k (2)^{1-k}  {\cal{Q}}_{e}}{k \vert\Sigma_{2k-1}\vert}\right]^{\frac{2k}{2k-1}},\qquad
p_r=-\mu,\qquad p_t=\frac{2^{k-2}}{r^{2k}}
\left[\frac{(-1)^k (2)^{1-k}  {\cal{Q}}_{e}}{k \vert\Sigma_{2k-1}\vert}\right]^{\frac{2k}{2k-1}}.
\end{eqnarray}}

\subsection{Five-dimensional AdS dyonic black holes and particular stealth configuration}
In five dimensions, the previous solution can be magnetically
charged in such a way that the magnetic charge also appears in the
metric function with a slower falloff of logarithmic order in
comparison to the mass. The corresponding model that sustains such
solution is given by the action (\ref{Lagdila}) by setting
$\phi=\psi_j=0$ and by considering $n=3$ extra gauge fields ${\cal
A}_{I}$ as well as the nonlinear Maxwell term with the exponent
$q=2$. In this case, the dyonic solution which can also be viewed as
an electric extension of the solution (\ref{magnsolution}) is given
by the ansatz (\ref{lineelementadm}) with
\begin{eqnarray}
\label{solutiondyonicfive} &&F(r)=-\frac{\Lambda}{6}r^2-\frac{2{\cal
M}}{3\vert\Sigma_3\vert r^2}-\frac{\ln r}{r^2} \left[\frac{{\cal
Q}_m^2}{\vert\Sigma_3\vert^2}-\frac{2\vert{\cal
Q}_e\vert^{\frac{4}{3}}}{\left(4\vert\Sigma_3\vert\right)^{\frac{4}{3}}}\right],\qquad N(r)=1,\nonumber\\
&&{\cal A}_1=\frac{{\cal Q}_m}{\vert\Sigma_3\vert}x_2\, dx_3,\quad
{\cal A}_2=\frac{{\cal Q}_m}{\vert\Sigma_3\vert} x_3\,dx_1,\quad
\quad {\cal A}_3=\frac{{\cal Q}_m}{\vert\Sigma_3\vert}
x_1\,dx_2,\qquad { A_0=-\frac{\vert{\cal
Q}_e\vert^{\frac{1}{3}}}{\left(4\vert\Sigma_3\vert\right)^{\frac{1}{3}}}\ln
r}.\nonumber
\end{eqnarray}
Before proceeding as before, we would like to point out that the
point defined by
\begin{eqnarray}
\vert {\cal Q}_e\vert=\frac{2^{\frac{5}{4}}\vert {\cal
Q}_m\vert^{\frac{3}{2}}}{\vert \Sigma_3\vert^{\frac{1}{2}}},
\label{stealthpoint}
\end{eqnarray}
is very special in the sense that the metric function reduces to the
Schwarzschild AdS metric with a flat horizon. This in turn implies
that the field equations at the point (\ref{stealthpoint}) can be
interpreted as a stealth configuration \cite{AyonBeato:2004ig}
defined on the Schwarzschild AdS background  since both side
(geometric and matter part) of the Einstein equations vanish separately, i. e.
\begin{eqnarray}
G_{\mu \nu}+\Lambda g_{\mu \nu} =&0&=
\sum_{I=1}^{3}\left({\cal{F}}_{(I)\mu \sigma} {\cal{F}}_{(I)
\nu}^{\phantom{\sigma \sigma \, } \sigma}\right)-\frac{1}{4}
{g}_{\mu \nu}\sum_{I=1}^{3}\left(
        {\cal{F}}_{(I) \sigma \rho}{\cal{F}}_{(I)}^{\sigma
\rho}\right)\\
&&+2
F_{\mu\sigma}F_{\nu}^{\,\,\sigma}\left(F_{\alpha\beta}F^{\alpha\beta}\right)-
\frac{1}{4}g_{\mu\nu}\left(F_{\alpha\beta}F^{\alpha\beta}\right)^2.\nonumber
\end{eqnarray}
Note that such stealth configuration but for a dyonic four-dimensional Reissner-Nordstrom
black hole was known in the case of an Abelian gauge field coupled to a particular Horndeski
term \cite{Babichev:2015rva} or for a generalized Proca field theory \cite{Cisterna:2016nwq}.

Outside the stealth point, the discriminant
associated to the zeros of the metric function $F$ is given by
\begin{eqnarray}
\Delta=\frac{2\Lambda}{3 A}e^{-\frac{8{\cal M}}{3\vert\Sigma_3\vert
A}},\qquad\mbox{with}\quad A=\frac{{\cal
Q}_m^2}{\vert\Sigma_3\vert^2}-\frac{2\vert{\cal
Q}_e\vert^{\frac{4}{3}}}{\left(4\vert\Sigma_3\vert\right)^{\frac{4}{3}}}.
\end{eqnarray}
Since, we are only considering the negative cosmological constant
case, we conclude that:
\begin{enumerate}[(i)]
\item For $A<0$, that is for
$$
\vert {\cal Q}_e\vert>\frac{2^{\frac{5}{4}}\vert {\cal
Q}_m\vert^{\frac{3}{2}}}{\vert \Sigma_3\vert^{\frac{1}{2}}},
$$
the solution has a single horizon located at
$$
r_h=e^{-\frac{W_{0}(\Delta)}{4}-\frac{2{\cal M}}{3\vert\Sigma_3\vert
A}},
$$
but  the solution does not satisfy the
dominant energy conditions neither the weak energy conditions since
the energy density $\mu=\frac{3A}{2r^4}$ is always negative.
\item For $A>0$, that is for
$$
\vert {\cal Q}_e\vert<\frac{2^{\frac{5}{4}}\vert {\cal
Q}_m\vert^{\frac{3}{2}}}{\vert \Sigma_3\vert^{\frac{1}{2}}},
$$
the solution represents a dyonic AdS black hole only if
$$
{\cal M} \geq \frac{3\vert\Sigma_3\vert
A}{8}\left[1-\ln\left(\frac{3A}{-2\Lambda}\right)\right],
$$
and in this case, the solution is shown to satisfy the dominant
energy conditions (\ref{DEC}).
\item Finally, for $A=0$ that is for
$$
\vert {\cal Q}_e\vert=\frac{2^{\frac{5}{4}}\vert {\cal
Q}_m\vert^{\frac{3}{2}}}{\vert \Sigma_3\vert^{\frac{1}{2}}},
$$
the solution represents a black hole stealth dyonic configuration on
the Schwarzschild AdS background where the horizon is located at
$$
r_h=\left(\frac{4{\cal
M}}{-\Lambda\vert\Sigma_3\vert}\right)^{\frac{1}{4}}.
$$
\end{enumerate}

\subsection{Purely magnetic Lifshitz black hole with dynamical
exponent $z=D-4$}
We now turn to derive examples with a different asymptotic behavior characterized by an
anisotropy scale between the time and the space, the so-called Lifshitz asymptotic. This anisotropy is
reflected by a dynamical exponent denoted usually by $z$
and defined such that the case $z=1$ corresponds to the AdS
isotropic case. Note that Lifshitz black holes have been insensitively studied during
the last decade, see for examples Refs. \cite{Lifs}

In order to obtain Lifshitz black holes, we consider the action
(\ref{Lagdila}) without axionic fields $\psi_j=0$ and with the
standard Maxwell term $q=1$. Note that the presence of the Maxwell
term is mandatory to ensure the Lifshitz asymptotic of the solution.
In even (resp. odd) dimension, the solution will be sustained by
$n=1$ (resp. $n=3$) extra gauge field(s) ${\cal A}_I$. In both case,
a purely magnetic Lifshitz black hole with dynamical exponent
$z=D-4$ is found through the ansatz (\ref{lineelementadm}) with
\begin{eqnarray}
&&F(r)=r^2-\frac{2{\cal M}}{(D-2)\vert\Sigma_{D-2}\vert
r^{2(D-4)}}-\frac{{\cal Q}_m^{2}}{\vert\Sigma_{D-2}\vert^2
r^{2(D-4)}} \ln(r),\qquad N(r)=r^{D-5},\nonumber\\
&&A_{0}=\sqrt {\frac{D-5}{2(D-3)}} {r}^{2(D-3)}, \quad e^{\phi}=
r^{\sqrt{{(D-2) (D-5)}}},\\
&&{\cal{A}}_{(1)}=\frac{\sqrt{2} {\cal Q}_m}{2\vert\Sigma_{D-2}\vert}
\sum_{i=1}^{\frac{D-2}{2}}\left(x_{2i-1} dx_{2i}-x_{2i} dx_{2i-1}\right), \mbox{ for even dimension},\nonumber\\
&&{\cal{A}}_{(I)}=\frac{{\cal
Q}_m}{2\vert\Sigma_{D-2}\vert}\sum_{J,K=1}^{3} \epsilon_{IJK} x_{J}
dx_{K}+ {\frac{\sqrt{6}}{6}}\frac{{\cal
Q}_m}{\vert\Sigma_{D-2}\vert} \sum_{i=2}^{\frac{D-3}{2}}\left(x_{2i}
dx_{2i+1}-x_{2i+1} dx_{2i}\right),\nonumber
\end{eqnarray}
for odd dimension with $I=\{1,2,3\}$. Here $\vert\Sigma_{D-2}\vert$
denotes the finite volume element of the compact $(D-2)-$dimensional
base manifold  and $\epsilon_{IJK}$ is defined as
\[ \epsilon_{IJK} = \left\{ \begin{array}{ll}
1 & \mbox{for any even permutation of $ (1,2,3),$}\\
-1 & \mbox{for any odd permutation of $ (1,2,3),$}\\
0 & \mbox{otherwise.}\\
\end{array} \right. \]
In this case, the coupling constants of the problem
must take the following form
$$
\Lambda=-(D-3)(2D-7),\quad \lambda=-2\sqrt{\frac{D-2}{D-5}},\quad
\alpha_1=\alpha_2=\alpha_3=-2\sqrt{\frac{D-5}{D-2}}.
$$
Proceeding as in the two previous examples, one notes that the
discriminant is always negative, and hence the existence of horizons
is again ensured provided that the mass satisfies the following
bound
\begin{eqnarray}
{\cal M}\geq \frac{(D-2){\cal
Q}_m^2}{4(D-3)\vert\Sigma_{D-2}\vert}\left[1-\ln\left(\frac{{\cal
Q}_m^2}{2(D-3)\vert\Sigma_{D-2}\vert^2}\right)\right].
\end{eqnarray}

For this lifshitz solution with dynamical exponent $z=D-4$, the energy density
 and the radial/tangential pressures are given by
\begin{eqnarray*}
&&\mu=(D-5)(D-3)+\frac{(D-2){\cal Q}_m^2}{2\vert\Sigma_{D-2}\vert^2 r^{2(D-3)}}+\frac{(D-2)(D-5)}{2r^2}F(r),\\
&& p_r=\mu-\frac{1}{\vert\Sigma_{D-2}\vert^2 r^{2(D-3)}}\left(2(D-3)(D-5)\vert\Sigma_{D-2}\vert^2 r^{2(D-3)}+(D-2){\cal Q}_m^2\right),\\
&& p_t=-\mu-\frac{1}{\vert\Sigma_{D-2}\vert^2 r^{2(D-3)}}\left(2(D-3)(D-5)\vert\Sigma_{D-2}\vert^2 r^{2(D-3)}+2{\cal Q}_m^2\right),
\end{eqnarray*}
and since
$$
p_r+\mu=\frac{(D-2)(D-5) F(r)}{r^2},\qquad p_t-\mu=-\frac{(D-2)(D-5) F(r)}{r^2}-\frac{(D-4){\cal Q}_m^2}
{\vert\Sigma_{D-2}\vert^2 r^{2(D-3)}},
$$
one can notice that the dominant energy conditions (\ref{DEC}) are satisfied outside the event horizon, that is for $F(r)\geq 0$.

\subsection{Axionic Lifshitz black hole with dynamical
exponent $z=D-2$}
We now consider the action (\ref{Lagdila})  with a source only given
by the axionic fields $\psi_j$ and with the standard Maxwell term
$q=1$ in order to sustain the Lifshitz asymptotic. In this case, an
axionic Lifshitz black hole solution with dynamical exponent $z=D-2$
is found to be
{\begin{eqnarray}
&&F(r)=r^2-\frac{2{\cal M}}{(D-2)\vert\Sigma_{D-2}\vert
r^{2(D-3)}}-\frac{{\cal Q}_a^{2}}{\vert\Sigma_{D-2}\vert^2
r^{2(D-3)}} \ln(r),\qquad N(r)=r^{D-3},\nonumber\\
&&A_{0}=\sqrt {\frac{D-3}{2(D-2)}} {r}^{2(D-2)}, \quad e^{\phi}=
r^{\sqrt{{(D-2) (D-3)}}},\\
&&\psi_{j}(x_{j})=-\frac{{\cal Q}_a}{\vert\Sigma_{D-2}\vert}
x_{j},\qquad \mbox{with } j=\{1,2,\cdots,D-2\},\nonumber
\end{eqnarray}}
where now ${\cal Q}_a$ denotes the axionic charge. The coupling
constants must be chosen such as
$$
\Lambda=-(D-2)(2D-5),\quad \lambda=-2\sqrt{\frac{D-2}{D-3}},\quad \beta_j=-2 \sqrt{\frac{D-3}{D-2}},\qquad
\mbox{with } j=\{1,2,\cdots,D-2\}.
$$
As for the previous case, the discriminant is negative and the mass
parameter must satisfy the following bound with respect to the
axionic charge in order to avoid naked singularity
{\begin{eqnarray}
{\cal M}\geq \frac{{\cal
Q}_a^2}{4 \vert\Sigma_{D-2}\vert}\left[1-\ln\left(\frac{{\cal
Q}_a^2}{2(D-2)\vert\Sigma_{D-2}\vert^2}\right)\right].
\end{eqnarray}}

As in the previous Lifshitz case, the dominant energy conditions (\ref{DEC}) are satisfied
outside the event horizon with
\begin{eqnarray*}
&&\mu=(D-2)(D-3)+\frac{(D-2){\cal Q}_a^2}{2\vert\Sigma_{D-2}\vert^2 r^{2(D-2)}}+\frac{(D-2)(D-3)}{2r^2}F(r),\\
&& p_r=\mu-\frac{1}{\vert\Sigma_{D-2}\vert^2 r^{2(D-2)}}\left(2(D-3)(D-2)\vert\Sigma_{D-2}\vert^2 r^{2(D-2)}+(D-2){\cal Q}_a^2\right),\\
&& p_t=-\mu-\frac{1}{\vert\Sigma_{D-2}\vert^2 r^{2(D-2)}}\left(2(D-3)(D-2)\vert\Sigma_{D-2}\vert^2 r^{2(D-2)}+{\cal Q}_a^2\right),\\
&& p_r+\mu=\frac{(D-2)(D-3) F(r)}{r^2},\qquad p_t-\mu=-\frac{(D-2)(D-3) F(r)}{r^2}-\frac{(D-3){\cal Q}_a^2}
{\vert\Sigma_{D-2}\vert^2 r^{2(D-2)}}.
\end{eqnarray*}

We now compute the DC conductivity $\sigma_{\tiny{\mbox{DC}}}$ of this solution which can
be expressed in term of the black hole horizon data
\cite{Donos:2013eha, Donos:2014cya} thanks to the  presence of the axionic fields homogenously distributed along the
coordinates of the planar base.  In order to achieve this task, we will follow
the prescriptions as given in these last references by first turning on the
following relevant perturbations\footnote{For simplicity, we only consider perturbations along
 one of the planar coordinate $x_1$.}
$$
\delta A_0=-Et+a_{x_1}(r),\qquad \delta g_{t{x_1}}=r^2 h_{t{x_1}}(r),\qquad \delta g_{r{x_1}}=r^2 h_{r{x_1}}(r),\quad \delta \psi_1=\chi_1(r),
$$
where $E$ is a constant. The  perturbed Maxwell current given by $J=\sqrt{-g}e^{\lambda\phi}F^{rx_1}$
is a conserved quantity along the radial coordinate. A straightforward computation along the same lines as those in
\cite{Donos:2013eha, Donos:2014cya} yields a DC conductivity $\sigma_{\tiny{\mbox{DC}}}$ given by
$$
\sigma_{\tiny{\mbox{DC}}}=\frac{\partial J}{\partial E}\Big{\vert}_{r_h}=r_h^{-D}+\frac{2(D-2)(D-3)\vert \Sigma_{D-2}\vert^2 r_h^{D-4}}{{\cal Q}_a^2}.
$$
As it should be expected in the absence of the axionic charge, the expression of the DC conductivity
$\sigma_{\tiny{\mbox{DC}}}$ will blows up.

\section{More general dyonic-axionic solutions in arbitrary
dimension}
The solutions derived previously can accommodate extra Noetherian
charges but in this case the location of the horizons is more
involved and can not be treated as before with the help of the
Lambert W functions. Nevertheless, for completeness, we report in
this section more general solutions, each of them having a dyonic
and an axionic charge. In order to achieve this task, we consider a
slightly different action than the one defined by Eq. (\ref{Lagdila}). Indeed, we will
add an extra Maxwell term without any nonlinearity $q=1$ since
it is known that electrically charged Lifshitz black holes require the
introduction of at least two Maxwell terms \cite{Tarrio:2011de}. We
then consider the following $D-$dimensional action
\begin{eqnarray*}
&&S[g_{\mu \nu},\phi, A_{(i) \mu}, {\cal{A}}_{(I)
\mu},\psi_{j}]=\int
d^Dx\,\sqrt{-g} \,{\cal L},\label{action}\\
&&{\cal L}=\frac{R-2 \Lambda}{2}-
\frac{1}{2}\partial_{\mu}\phi\partial^{\mu}\phi-\frac{1}{4}\,\sum_{i=1}^{2}
e^{\lambda_{i}\phi}F_{(i) \mu \nu}F_{(i)}^{\mu \nu}-
 e^{\alpha\phi}\left(\frac{1}{4}\sum_{I=1}^{n} {\cal{F}}_{(I) \mu \nu}{\cal{F}}_{(I)}^{\mu
\nu}+\frac{1}{2}\sum_{j=1}^{D-2} \partial_{\mu} \psi_{j}
\partial^{\mu} \psi_{j}\right),
\end{eqnarray*}
with $F_{(i)\mu \nu}=\partial_{\mu} A_{(i)\nu}-\partial_{\nu}
A_{(i)\mu}$ for   $i=\{1,2\}$ and $n$ extra gauge fields that will sustain the magnetic charge, ${\cal{F}}_{(I)\mu
\nu}=\partial_{\mu} {\cal{A}}_{(I)\nu}-\partial_{\nu}
{\cal{A}}_{(I)\mu}$ for $I=1,\cdots n$ with $n=1$ in even dimension and $n=3$ in odd
dimension. The field equations read
\begin{subequations}
    \begin{eqnarray*}
        G_{\mu \nu}+\Lambda g_{\mu \nu}=\Big(\nabla_{\mu} \phi
        \nabla_{\nu} \phi-\frac{1}{2} g_{\mu \nu}
        \nabla_{\sigma}\phi \nabla^{\sigma}\phi\Big)+\sum_{i=1}^{2}
        \Big(e^{\lambda_{i} \phi}
        F_{(i)\mu \sigma} F_{(i) \nu}^{\phantom{\sigma \sigma} \sigma}
        -\frac{1}{4} g_{\mu \nu}
        e^{\lambda_{i}
        \phi} F_{(i) \sigma \rho}F_{(i)}^{\sigma \rho}\Big)&&\nonumber\\
        e^{\alpha \phi}\sum_{j=1}^{D-2}\Big(\partial_{\mu} \psi_{j}
        \partial_{\nu} \psi_{j}-\frac{1}{2} g_{\mu \nu}
        \partial_{\sigma}\psi_{j} \partial^{\sigma}\psi_{j}\Big)-\frac{1}{4} {g}_{\mu \nu} e^{\alpha \phi}\sum_{I=1}^{n}
{\cal{F}}_{(I) \sigma \rho}{\cal{F}}_{(I)}^{\sigma
\rho}+e^{\alpha \phi}\sum_{I=1}^{n}{\cal{F}}_{(I)\mu \sigma} {\cal{F}}_{(I) \nu}^{\phantom{\sigma \sigma } \sigma},\label{eqmotiong}&&\\
       \nabla_{\mu} \left(e^{\lambda_{i} \phi} F_{(i)}^{\mu
        \nu}\right)=0,\label{eqmotionAmu}&&\\
        {\nabla}_{\mu} \left(e^{\alpha \phi} {\cal{F}}_{(I)}^{\mu \nu}\right)=0, \label{eqmotionAxi}&&
        \\
        \Box \psi_{j}=0,\label{eqmotionaxionic}&&
        \\
        \Box \phi- \sum_{i=1}^{2}\left(\frac{\lambda_{i}}{4} e^{\lambda_{i} \phi}
         F_{(i) \sigma \rho}
        F_{(i)}^{\sigma
        \rho}\right)-\alpha e^{\alpha\phi}\left(\frac{1}{4}\sum_{I=1}^{n} {\cal{F}}_{(I) \mu \nu}{\cal{F}}_{(I)}^{\mu
\nu}+\frac{1}{2}\sum_{j=1}^{D-2} \partial_{\mu} \psi_{j}
\partial^{\mu} \psi_{j}\right)=0,\label{eqmotionphi}&&
    \end{eqnarray*}
\end{subequations}
A general Lifshitz dyonic-axionic solution with arbitrary dynamical
exponent $z$ of the field equations is given by
\begin{eqnarray}
F(r)=r^2-\frac{\lambda^2}{(D-2-z) r^{2z-2}}-{\frac
{M}{{r}^{z+D-4}}}+{\frac {Q^{2}}{ (D-2)(z+D-4) {r}^{2
(z+D-4)}}}-\frac{ P^2}{(D-4-z) r^{2 z}} , \label{F}
\end{eqnarray}
together with
\begin{eqnarray}
\label{solz} N(r)&=&r^{z-1},\quad A_{(1)t}=\sqrt {\frac{z-1}{z+D-2}}
{r}^{z+D-2}
dt,\quad A_{(2)t}=-\frac{Q}{(z+D-4) r^{z+D-4}}dt, \nonumber\\
\psi_{j}(x_{j})&=&\lambda x_{j},\qquad e^{\phi}= r^{\sqrt{{(D-2)
(z-1)}}}, \\
{\cal{A}}_{(1)}&=&\frac{\sqrt{2} P}{2} \sum_{i=1}^{\frac{D-2}{2}}\left(x_{2i-1} dx_{2i}-x_{2i} dx_{2i-1}\right), \mbox{ in even dimensions},\nonumber\\
{\cal{A}}_{(I)}&=&\frac{P}{2}\sum_{J,K=1}^{3} \epsilon_{IJK} x_{J}
dx_{K}+ {\frac{\sqrt{6}}{6}}P \sum_{i=2}^{\frac{D-3}{2}}\left(x_{2i}
dx_{2i+1}-x_{2i+1} dx_{2i}\right),\mbox{ in odd dimensions}\nonumber
\end{eqnarray}
provided that the coupling constants are tied as follows
\begin{eqnarray*}\label{couplingconstants}
\Lambda&=&-\frac{(z+D-2) (z+D-3)}{2},\quad
\lambda_{1}=-2\sqrt{\left(\frac{D-2}{z-1}\right)}, \quad
\lambda_2=-\alpha=2\sqrt{\left(\frac{z-1}{D-2}\right)}.
\end{eqnarray*}
Note that in the AdS limit $z=1$, the dilaton field $\phi$
disappears as well as the Maxwell potential $A_{(1)t}$ which is
precisely responsible to sustain the Lifshitz asymptotic $z\not=1$.
The thermodynamical variables of the solution can be computed using
the Hamiltonian formalism \cite{Gibbons:1976ue} yielding
\begin{eqnarray}
{\cal{M}}&=&\frac{1}{2} (D-2) M \vert\Sigma_{D-2}\vert,\nonumber\\
{\cal{S}}&=&2 \pi r_{h}^{D-2}\vert\Sigma_{D-2}\vert,
\end{eqnarray}
with the electric, magnetic and axionic potentials and charges
\begin{eqnarray}
{\cal{Q}}_{e}&=&Q \vert\Sigma_{D-2}\vert,\qquad \Phi_{e}=\left(\frac{ r_{h}^{4-D-z}}{z-4+D}\right)\,Q ,\nonumber\\
{\cal{Q}}_{m}&=& \vert\Sigma_{D-2}\vert P,\qquad
\Phi_{m}=-\left({\frac { D-2 }{D-4-z}}\right) r_{h}^{D-4-
z}\,P,\\
{\cal{Q}}_{j}&=&-\vert\Sigma_{D-2}\vert\lambda,\qquad
\hat{\Psi}_{j}(r_h)={\frac { r_{h}^{D-2-z}\,\lambda}{D-2-z}} ,
\mbox{ with } j=\{1,2,\cdots, D-2\}.\nonumber
\end{eqnarray}
It is a simple exercise to check the consistency of the first
law\begin{eqnarray}\label{firstlaw} d{\cal{M}}=T
d{\cal{S}}+\Phi_{e}d{\cal{Q}}_{e}+\Phi_{m}d{\cal{Q}}_{m}
+\sum_{j=1}^{D-2}\hat{\Psi}_{j}(r_h) d{\cal{Q}}_{j},
\end{eqnarray}
where the temperature is given by
\begin{equation}
T=\frac{N(r)F'(r)}{4 \pi} \Big{|}_{r=r_{h}}=\frac{1}{4 \pi}
\left[(D-2+z) r_{h}^{z}-\frac{Q^{2}}{(D-2)r_{h}^{z+2D-6}}
-\frac{P^{2}}{r_{h}^{z+2}}-\frac{\lambda^{2}}{r_{h}^{z}}\right].
\end{equation}

It is clear that from the expression of the metric function, the
cases $z=D-4$ and $z=D-2$ must be treated separately. In fact, for
$z=D-4$, one yields a metric function involving a logarithmic
magnetic contribution
\begin{equation}
F(r)=r^2-\frac{M}{r^{2(D-4)}}+\frac{Q^{2}}{2 (D-2) (D-4)
r^{4(D-4)}}-\frac{P^{2}}{r^{2(D-4)}} \ln(r)-\frac{\lambda^2}{2
r^{2(D-5)}},
\end{equation}
and the remaining fields are given by (\ref{solz}) with $z=D-4$. The
thermodynamical quantities computed by means of the Euclidean method
\cite{Gibbons:1976ue} read
\begin{eqnarray}
{\cal{M}}&=&\frac{1}{2} (D-2) M \vert\Sigma_{D-2}\vert,\nonumber\\
T&=& \frac{1}{4 \pi} \left[2(D-3)
r_{h}^{D-4}-\frac{Q^{2}}{(D-2)r_{h}^{3D-10}}
-\frac{P^{2}}{r_{h}^{D-2}}-\frac{\lambda^2}{r_{h}^{D-4}}\right] ,\quad {\cal{S}}=2 \pi r_{h}^{D-2}\vert\Sigma_{D-2}\vert,\nonumber\\ \nonumber\\
{\cal{Q}}_{e}&=&Q \vert\Sigma_{D-2}\vert,\qquad \Phi_{e}=\frac{ Q}{2(D-4) r_{h}^{2(D-4)}} ,\\
{\cal{Q}}_{m}&=&\vert\Sigma_{D-2}\vert P,\qquad \Phi_{m}=-(D-2) \ln(r_{h})\,P,\nonumber\\
{\cal{Q}}_{j}&=&-\vert\Sigma_{D-2}\vert\lambda,\qquad
\hat{\Psi}_{j}(r_h)={\frac { r_{h}^{2}\,\lambda}{2}} , \mbox{ con }
j=\{1,2,\cdots, D-2\}.\nonumber
\end{eqnarray}
Finally, the solution with $z=D-2$ is given by
\begin{equation}
F(r)=r^2-\frac{M}{r^{2(D-3)}}+\frac{Q^{2}}{2 (D-2) (D-3)
r^{4(D-3)}}+\frac{P^{2}}{2 r^{2(D-4)}}-\frac{ \lambda^{2}
\ln(r)}{r^{2(D-3)}}.
\end{equation}
with (\ref{solz}), and the thermodynamic parameters associated to
the solution are
\begin{eqnarray}
{\cal{M}}&=&\frac{1}{2} (D-2) M \vert\Sigma_{D-2}\vert, \nonumber\\
T&=& \frac{1}{4 \pi} \left[2(D-2)
r_{h}^{D-2}-\frac{Q^{2}}{(D-2)r_{h}^{3D-8}}
-\frac{P^{2}}{r_{h}^{D}}-\frac{\lambda^2}{r_h^{D-2}}\right] ,\quad {\cal{S}}=2 \pi r_{h}^{D-2}\vert\Sigma_{D-2}\vert,\nonumber\\ \nonumber\\
{\cal{Q}}_{e}&=&Q \vert\Sigma_{D-2}\vert,\qquad \Phi_{e}=\frac{ Q}{2(D-3) r_h^{2(D-3)}} ,\\
{\cal{Q}}_{m}&=&\vert\Sigma_{D-2}\vert P,\qquad \Phi_{m}= \frac{(D-2)\,P}{2 r_h^2},\nonumber\\
{\cal{Q}}_{j}&=&-\vert\Sigma_{D-2}\vert\lambda,\qquad
\hat{\Psi}_{j}(r_h)=\lambda \ln(r_h) , \mbox{ con } j=\{1,2,\cdots,
D-2\}.\nonumber
\end{eqnarray}
In both cases, that is for $z=D-4$ and $z=D-2$, it is easy to see that the first
law (\ref{firstlaw}) holds.

\section{Conclusion}

Here, we have presented a dyonic extension of the five-dimensional
Boulware-Deser solution for the Einstein-Gauss-Bonnet theory. The
emergence of a magnetic charge is shown to be possible for a flat
horizon and by considering at least three different Maxwell
invariants. The magnetic contribution in the metric function has a
logarithmic falloff but still yields to finite physical quantities.
As usual, one of the two branches has a well-defined GR-limit with a
magnetic logarithmic falloff term. For suitable bounds between the
mass and the magnetic charge, the purely magnetic GR solution
can be shown to admit an inner and outer horizons. These latter are
given in terms of the two real branches of the Lambert W functions. We have noticed
that this bound's mass was due to the fact
that the magnetic charge in the metric has a slower falloff of
logarithmic order than the mass. Exploiting this observation, we
have  derived other examples of solutions sharing these same
properties for different models and different asymptotics. For
example, we have obtained an electrically charged AdS black hole
solution in odd dimension for a nonlinear Maxwell theory  with a
single horizon in dimensions $D=5\,\,\mbox{mod}\, 4$ and with two
horizons in $D=3\,\,\mbox{mod}\, 4$. Interestingly enough, a dyonic
configuration with logarithmic falloff of the electric and magnetic
charges was also derived in five dimensions. Depending on the
strength of the electric charge with respect to the magnetic charge,
the solution can have one or two horizons, and in this latter case,
the mass must satisfy a certain bound. Moreover, for a precise
relation between the electric and the magnetic charges, the solution
turns out to be a stealth dyonic configuration defined on the
Schwarzschild AdS background. For the asymptotic AdS solutions, we have remarked
that our black hole solutions presenting an inner and outer horizons always
satisfy the dominant energy conditions (\ref{DEC}) while these conditions
even in their weak version do not hold for our solutions with a single horizon. This can be explained
by the fact that the metric functions in our set-up were of the following form
\begin{eqnarray}
F(r)=-\frac{2\Lambda r^2}{(D-1)(D-2)}-\frac{2{\cal M}}{(D-2)\vert\Sigma_{D-2}\vert r^{D-3}}-\frac{{\cal N}\ln r}{r^{D-3}},
\label{F}
\end{eqnarray}
where ${\cal N}$ represent the additional Noetherian charge with slower falloff of logarithmic order
than the mass ${\cal M}$. The corresponding discriminant associated to the zeros of the metric function $F$
(\ref{discriminant}) is given by
$$
\Delta=\frac{2\Lambda }{(D-2){\cal N}}\,\,e^{-\frac{2(D-1){\cal M}}{(D-2)\vert\Sigma_{D-2}\vert{\cal N}}}.
$$
Now, since we are considering the AdS case, it is clear that for ${\cal N}<0$, the discriminant will be
positive and, hence the solution will represent a black hole
with a single horizon for any value of the mass ${\cal M}$. On the other hand, for ${\cal N}>0$, one has $\Delta<0$ and consequently
the solution will be a black hole provided that the mass satisfies the following bound
$$
{\cal M}> \frac{(D-2)\vert\Sigma_{D-2}\vert{\cal N}}{2(D-1)}\left[1-\ln\left(\frac{(D-2){\cal N}}{-2\Lambda}\right)\right],
$$
and in this case, the solution presents two horizons. On the other hand, the energy density, the radial
and tangential pressures are given generically by
$$
\mu=\frac{(D-2){\cal N}}{2\,r^{D-1}},\qquad p_r=-\mu,\qquad p_t=\frac{{\cal N}}{2\,r^{D-1}},
$$
and hence it is evident that the dominant energy conditions (\ref{DEC}) will only be satisfied for the solutions
with ${\cal N}>0$. It seems to be physically acceptable that solutions without any restrictions on the mass
do not satisfy the dominant or the weak or even the null energy conditions. On the other hand, our examples of
black holes with a bound's mass verify the dominant energy conditions. It will be interesting to explore more deeply
this relation between the lack of restriction on the mass with the absence of energy conditions.

We also mention that a necessary condition to obtain AdS black holes with an Ansatz of the form
$$
ds^2=-F(r)dt^2+\frac{dr^2}{F(r)}+r^2(dx_1^2+\cdots dx_{D-2}^2),
$$
with a metric function given by (\ref{F}) is that the energy momentum tensor of the matter source $T_{\mu\nu}$
satisfies $T_t^t+(D-2)T_i^i=0$ without summation for the planar indices $i$. Indeed, in this case, the consistency
of the Einstein equations $T_t^t+(D-2)T_i^i+(D-1)\Lambda=0$ yields to a nonhomogeneous Euler's differential
equation of second-order
$$
\frac{(D-2)F^{\prime\prime}}{2}+\frac{(D-2)(2D-5)F^{\prime}}{2r}+\frac{(D-2)(D-3)^2 F}{2r^2}=-(D-1)\Lambda,
$$
whose characteristic polynomial has a double root given by $r^{-D+3}$ and hence the general solution of this Euler's
equation is given by Eq. (\ref{F}).

We have also presented two other examples with Lifshitz
asymptotics with fixed values of the dynamical exponent with a magnetic charge and an axionic charge. The emergence
of such asymptotic solutions is essentially due to the presence of dilatonic fields. Note that there also exist
Lifshitz black holes with a logarithmic falloff in the case of higher-order gravity \cite{AyonBeato:2010tm}. Finally, for completeness, we
have extended the previous solutions to accommodate a dyonic as well
an axionic charge in arbitrary dimension.

\section{Appendix: The Lambert W functions}
The Lambert W functions are a set of functions that represent the
countably infinite number of solutions denoted by $W_k(z)$ of the
equation
\begin{eqnarray}
W e^W=z, \label{eqLW}
\end{eqnarray}
for a given $z\in\mathds{C}$. There are only two real-valued
branches of the Lambert W functions that are denoted, by convention,
$W_0$ and $W_{-1}$ with  $W_0: [-\frac{1}{e},\infty[\, \rightarrow\,
[-1,\infty[$ and $W_{-1}: [-\frac{1}{e},0[ \,\rightarrow \,]-\infty,
-1[$ with the convention that
$W_0(-\frac{1}{e})=W_{-1}(-\frac{1}{e})=-1$. The Lambert W functions
appear for the resolution of the equations of the form
\begin{eqnarray}
a x+b\ln (x)+c=0,\qquad a\not=0,\quad b\not=0. \label{eqLW1}
\end{eqnarray}
Indeed, by defining $w=\ln (x)$, the equation (\ref{eqLW1}) becomes
$a e^w+bw+c=0$, which is equivalent after some basic algebraic
manipulations to (\ref{eqLW}) with $W=-w-\frac{c}{b}$ and
$z=\Delta$, where the discriminant is defined by
\begin{eqnarray}
\Delta=\frac{a}{b}e^{-\frac{c}{b}}. \label{discriminant}
\end{eqnarray}
It is then clear that
\begin{enumerate}[(i)]
\item  If $\Delta\geq 0$ or $\Delta=-\frac{1}{e}$, the equation
(\ref{eqLW1}) admits a unique solution in $\mathds{R}$ given by
\begin{eqnarray}
x=e^{-W_0(\Delta)- \frac{c}{b}}. \label{Deltapos}
\end{eqnarray}
\item  If $\Delta \in ]-\frac{1}{e}, 0[$, the equation
(\ref{eqLW1}) has two real solutions given by
\begin{eqnarray}
x_1=e^{-W_0(\Delta)- \frac{c}{b}},\qquad x_2=e^{-W_{-1}(\Delta)-
\frac{c}{b}}. \label{Deltaneg}
\end{eqnarray}
\item Finally, if $\Delta<-\frac{1}{e}$, the equation
(\ref{eqLW1}) does not admit real roots.
 \end{enumerate}

\vspace*{2cm} {\bf Acknowledgments:} {
 MB is supported by grant Conicyt/ Programa Fondecyt de Iniciaci\'on en Investigaci\'on No. 11170037.}


\end{document}